%% file: IEEE_paper_main.tex
\documentclass[%
	final,
	journal, %conference,% conference, journal, technote, peerreview, peerreviewca
    comsoc,%comsoc, compsoc, transmag
	letterpaper,% letterpaper, a4paper, cspaper
	oneside,% oneside, twoside
	twocolumn,% onecolumn, twocolumn % IEEE uses always twocolumn % onecolumn just for info.
	nofonttune,%
]{IEEEtran}%
\input{./organization/preamble.tex}%
\input{./organization/makros.tex}%
\input{./organization/settings.tex}%
\usepackage{newtxmath}
\usetikzlibrary{external}
\makeatletter
\let\blx@rerun@biber\relax
\makeatother

\pgfplotsset{
  grid style = {
   line width = 0.1pt
  }
}
				\newcommand{\disablewr}[1]{#1}%
				\newcommand{\newcommanddisw}[3]{\newcommand{#1}[1]{\disablewr{\textcolor{#2}{#3}}}}%
\renewcommand{\disablewr}[1]{}%
\definecolor{todocol}{named}{red}
\newcommanddisw{\todo}{todocol}{ToDo: #1}%
\definecolor{migucol}{named}{purple}%
\newcommanddisw{\migucom}{migucol}{{@}comment: #1}%
\newcommanddisw{\miguhigh}{migucol}{#1}%
	\newcommand{\TempDisplayPreparation}{\disablewr{%
		\section{Draft-State: Comment Color Code}\noindent%
		\todo{Comments: ToDos}\nl%
		\migucom{Comments: Michael Gundall}\nl%

	}}%
\begin{document}%
%Design/Introduction of a Docker-based Virtual Process Controller that enables dynamical reprogramming/reconfiguration%
\title{%
Introduction of an Architecture for Flexible Future Process Control Systems as Enabler for Industry 4.0
\thanks{This research was supported by the German Federal Ministry of Education and Research (BMBF) within the project \gls{tacnet4.0} under grant number 16KIS0712K. The responsibility for this publication lies with the authors. This is a preprint of a work accepted but not yet published at the IEEE 25rd International Conference on Emerging Technologies and Factory Automation (ETFA). Please cite as: M. Gundall, C. Glas, and H.D. Schotten: “Introduction of an Architecture for Flexible Future Process Control Systems as Enabler for Industry 4.0”. In: 2020 IEEE 25rd International Conference on Emerging Technologies and Factory Automation (ETFA), IEEE, 2020.}
}%
%
%
%\input{./organization/IEEE_authors.tex}%
\input{./organization/IEEE_authors-long.tex}%
\maketitle
%
%For reaching the aforementioned Industrie 4.0 goals a flexible production is seen as the key. For reaching flexibility 
%\glspl{icps} forming the \gls{iiot} serve as the basis for a smart manufacturing. However, in recent industry deployments, legacy equipment exists, that does not allow any kind of flexibility or integration of icps. In this paper we provide a architecture, that allows a flexible reconfiguration of future process control systems. The architecture design follows the 4+1 model approach as you can find in literature and therefore gives an holistic architecture. Furthermore we give insights in recent challenges and outline future development steps.
%Dass \gls{iiot}, das im wesentlichen aus \glspl{icps} besteht, ermöglicht viele neue Use Cases, die eine intelligente Fertigung ermöglichen. In industriellen Bestandsanlagen existieren jedoch überwiegend Altgeräte und Technologien, die dieses Maß an Flexibilität nicht erlauben. In diesem Beitrag stellen wir eine Architektur vor, die eine flexible Rekonfiguration zukünftiger Prozessleitsysteme ermöglicht. Dabei wurde außerdem auf eine hohe Verfügbarkeit und Ausfallsicherheit des Systems geachtet, wie es im industriellen Umfeld benötigt wird. Das Architekturdesign folgt dem 4+1-Modellansatz, wie er in der Literatur zu finden ist. Somit ist gewährleistet, dass das Ergebnis eine ganzheitliche Architektur ist. Darüber hinaus geben wir Einblicke in aktuelle Herausforderungen und skizzieren die zukünftigen Entwicklungsschritte.
\begin{abstract}%
%The \acrlong{iiot}, which basically consists of \acrlong{icps}s, allows the realization of numerous novel use cases that are key enabler for a smart manufacturing. 
The term Industry 4.0, which refers to the fourth industrial revolution, aims at the digitalization of industries, including all kinds of production assets. With the help of these, so-called “\acrlong{icps}s", which form the \acrlong{iiot}, numerous novel use cases that are key enabler for a smart manufacturing, can be realized. However, existing facilities mainly consist of legacy equipment and technologies that do not offer these kind of flexibility. To address this issue, we introduce an architecture that allows a flexible reconfiguration and redeployment of future process control systems. Moreover, a high system availability and reliability, as required by industrial applications, has been taken into account. The architectural design follows the 4+1 model approach as it is available in the literature. This ensures, that the design results in a holistic architecture. Additionally, we provide insights into first results and outline future development steps.

\end{abstract}%
% \fi
\IEEEpeerreviewmaketitle
%
%
%
%
%
% Examples: floats, figure, subfigure (two column float), tables
% -------------------------------------------------------------------
% \input{./organization/backing/IEEETemplateProvidedExamples.tex}
% -------------------------------------------------------------------
% Note that the IEEE does not put floats in the very first column
% - or typically anywhere on the first page for that matter. Also,
% in-text middle ("here") positioning is typically not used, but it
% is allowed and encouraged for Computer Society conferences (but
% not Computer Society journals). Most IEEE journals/conferences use
% top floats exclusively. 
% Note that, LaTeX2e, unlike IEEE journals/conferences, places
% footnotes above bottom floats. This can be corrected via the
% \fnbelowfloat command of the stfloats package.
%
%
%“ ”
%
%#################################################################
%##=========================================================######
%##---------------------------------------------------------######
%#################################################################
%##=========================================================######
%##---------------------------------------------------------######
\section{Introduction}%
\label{sec:Introduction}
%##=========================================================######
%#################################################################
A highly flexible manufacturing is seen as one of the most relevant scenarios for factories of the future. Accordingly, the reconfiguration or even redeployment of process controllers in very short intervals, e.g. before each workpiece, is conceivable. In industrial environments, typically %distributed systems consisting of several hardware controllers, such like 
\glspl{plc} are used \cite{6246692}. These  controllers are highly sophisticated for continuous control tasks, but typically don't provide any kind of flexibility and have to be stopped, e.g. for deploying an update of the program logic. Hence, they are not suitable to solve this task.
 
To overcome this issue, the virtualization of process controllers or process control functions is a suitable approach. The virtualization and the application of cloud services are well-known topics in the \gls{it}, but these days, virtualization concepts are also investigated for industrial applications. %, belonging to the \gls{ot}.
Therefore, the authors in \cite {6837587} carried out a case study for a PC-based virtualized \gls{plc}, which was realized on the basis of a \gls{vm}. The results show that while its implementation is applicable to soft real-time use cases, it lacks hard real-time constraints. With the emerging container-based virtualization, a promising technology is available that is capable of increasing performance. Consequently, a performance comparison between virtual machines and container virtualization was conducted \cite{7095802}. In particular Docker container and \glspl{kvm} were compared. The benchmarks showed that the performance of Docker containers is slightly worse than the performance of native \gls{os}, but better compared to the use of \gls{kvm}. In addition, \cite{10.1145/2851613.2851737,goldschmidt2018container,indin2020} analyzed the use of container technology for the virtualization of industrial automation systems in terms of determinism, \gls{rt} capabilities, performance, and security. For this reason \cite{goldschmidt2018container, 8502526} propose architectures for flexible industrial control systems, but limit them to container-based and IEC~61499-based controllers. Based on the findings of \cite{indin2020}, which compare the use of bare-metal, VMs, and containers as platform for the virtualization of industrial automation systems, we decided to develop a novel architecture using the 4+1 model \cite{469759}. This results, among other things, in a logical view that is technology independent and can be mapped to each of the technologies. Thus, the main contributions of this paper are:
\textbf{
\begin{itemize}
    \item Introduction of the developed architecture that allows a flexible production, based on most important industrial use cases and their requirements.
    \item Virtualization strategy that is able to serve as basis for the realization of our concept and its building blocks.
\end{itemize}
}

Therefore, the paper is structured as follows: Sec. \ref{sec:Background} describes requirements that serve as basis for our architecture, while the architecture will be detailed in Sec. \ref{sec:Architectural Design}. In addition, Sec. \ref{sec: Virtualization Strategy} gives insights about the planned virtualization strategy as well as first results. Finally, Sec. \ref{sec:Conclusion} concludes the paper.

%#################################################################
%##=========================================================######
%##---------------------------------------------------------######
\section{Functional Requirements}%
\label{sec:Background}
%##=========================================================######
%#################################################################
To fulfill industrial use cases and thus realizing a smart manufacturing, \cite{goldschmidt2018container} identified several functional requirements that must be supported by virtualized industrial automation systems, that have been extended by \cite{8502526}. The most relevant features for our investigations are explained below.

\subsubsection{Reconfiguration} 
As already mentioned a high flexibility is required. Therefore, both the firmware of the controller and the user-defined control program have to be updated during operation. This means that the reconfiguration process has to be performed without a downtime of the system.%production plant. 

\subsubsection{Redeployment}
The redeployment process is a special case of reconfiguration. In this case, the virtualized industrial automation system must be redeployed during normal operation on another hardware node. This feature is required by mobile devices that may change their location during operation (e.g. change of the factory hall). Furthermore, the case that a system component requires an update or maintenance and is temporarily unavailable is also covered by this feature.

%\subsubsection{Interruption-free Hardware Replacement}

\subsubsection{Resilience and Self-Healing}
Very characteristic for industrial applications are the high demands on availability, which are very different from applications on the office floor. Industrial applications, belonging to the use case group of closed loop motion control, allow only a maximum failure of one minute per year \cite{8502649}. Since this availability cannot be guaranteed by the equipment of the office floor as a rule, redundancy measures should be taken, whereby the failure of a redundant controller should not affect the process. Since the required redundancy is no longer given after the failure of a system component, an automatic and seamless start of a further redundant instance on a separate hardware node should be triggered. 

%#################################################################
%##=========================================================######
%##---------------------------------------------------------######
\section{Architecture for Novel Virtualized industrial automation systems}%
\label{sec:Architectural Design}
%##=========================================================######
%#################################################################
This section introduces an architecture that is capable of addressing the needs discussed in the preceding section. Thereby, the design of this architecture should comply with the rules of the 4+1 architectural model \cite{469759}. This guideline consists of “4" so-called views (logical view, process view, physical view, and development view) that are used to describe an comprehensive architecture. These views are completed by the scenarios, which are indicated by the “+1", and describe the use case or more fine-grained processes of an use case for which the corresponding architecture part is being developed. In the next step, the resulting architecture parts form the overall architecture. 

\begin{figure*}[htbp]
%\centerline{\includegraphics[width=0.7\textwidth, trim = 220 160 270 130, clip]{figures/LogicalView.pdf}}
\centerline{\includegraphics[width=0.7\textwidth, trim = 220 160 270 130, clip]{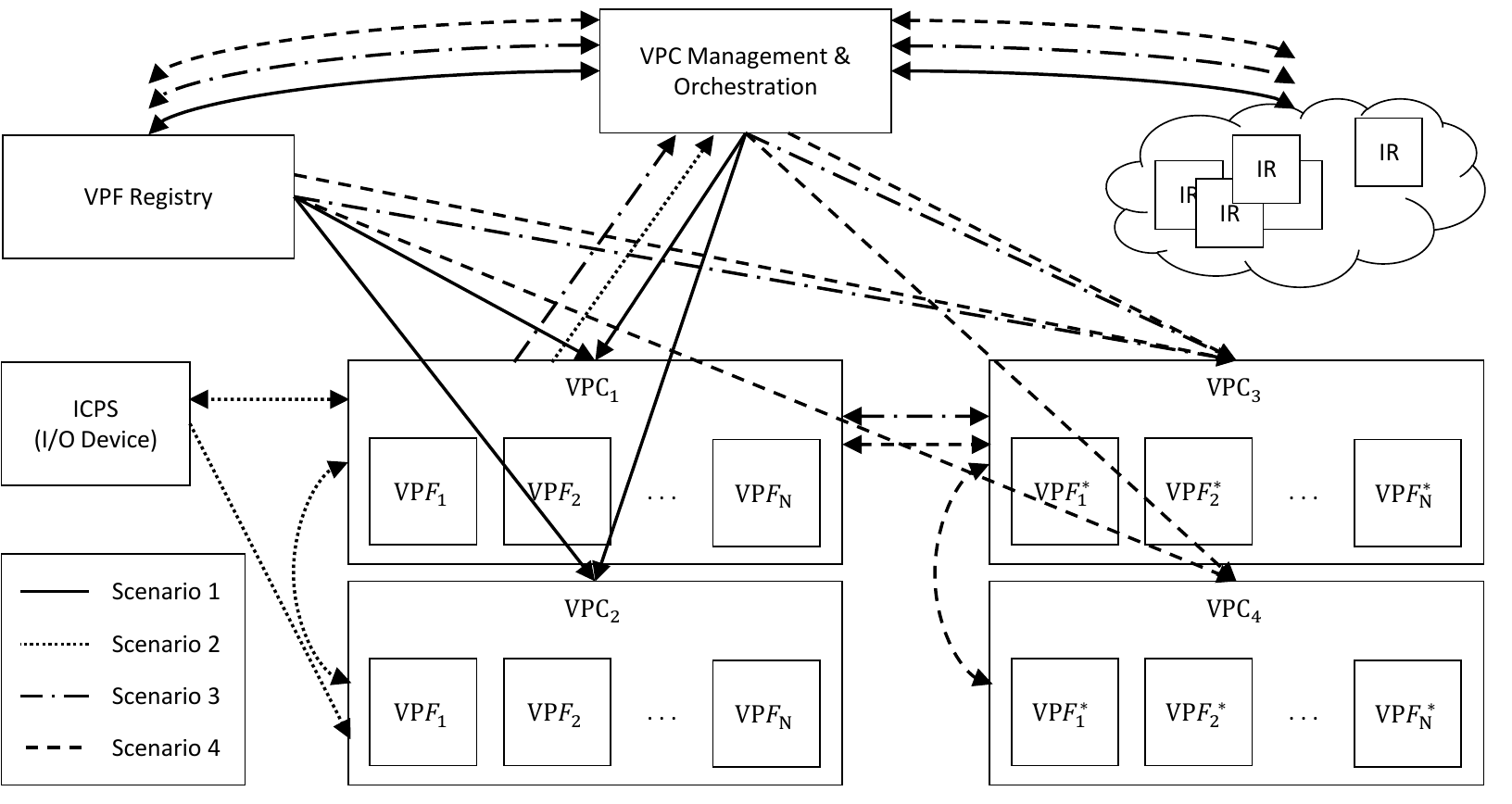}}
\caption{Technology independent logical view of the proposed architecture}
\label{fig:Logical view of the proposed architecture}
\end{figure*}

With the help of the logical view, the functional dependencies of entities and their interactions are shown for the specific scenario. After this procedure has been applied to each scenario mentioned above, all dependencies can be represented in a joint figure. The resulting architecture, which is shown in Fig.~\ref{fig:Logical view of the proposed architecture} is explained below using the four scenarios on which the architecture was designed. 

%In the following, the resulting architecture, which is shown in Fig. \ref{fig:Logical view of the proposed architecture}, is explained based on each on the four scenarios on which the architecture has been build. 

%As already mentioned, each scenario describes a specific process of which should be covered by the architecture. In the initial phase, there are four main scenarios that should be considered:

%\begin{itemize}
 %   \item \textbf{Start up (Scenario 1):} The start up scenario specifies the procedure for the bootstrapping of the system and its components. 
%    \item \textbf{Normal operation (Scenario 2):} This scenario describes the state where the system is running without faults and no events such as a reconfiguration or redeployment being recently triggered.
%    \item \textbf{Replacement (Scenario 3):} Here the procedure for replacing a failed \gls{vpc} is presented to ensure the resilience of the system.  
%    \item \textbf{Reconfiguration (Scenario 4):} In order to support smart manufacturing use cases, this scenario covers the reconfiguration or redeployment of \glspl{vpc}. 
%\end{itemize}

\subsection{Start up (Scenario 1)}
The start up scenario specifies the procedure for the bootstrapping of the system components. During this process, the \gls{vpcmo} searches for all available \glspl{ir}. An \gls{ir} can be specified as a component that is in an inactive state but indicates that it has available resources ready for deployment. To keep this information updated, the \gls{vpcmo} cyclically sends discovery messages, which can also serve as keep-alive message to indicate a failed \gls{ir}. When new \glspl{ir} are available, they register to the \gls{vpcmo}. If a \gls{vpc} deployment is requested from the \gls{vpcmo}, it determines the most appropriate \gls{ir}, %promotes it to the active \gls{vpc} 
sets it active ($VPC_1$) and transfers the configuration data to it. Additionally, the \glspl{vpf} that are executed by the \gls{vpc} are downloaded from the \gls{vpf} Registry. The specific \glspl{vpf} can be executed either cyclically or acyclically, and their complexity can vary depending on each application and use case.  Based on the required availability, this procedure is repeated for a defined number of inactive \glspl{vpc} that serve as a backup. In this case one inactive \gls{vpc} is assumed ($VPC_2$).

\subsection{Normal operation (Scenario 2)}
This scenario describes the state in which the system runs without errors and no events, such as reconfiguration or redeployment, occurred. The \gls{vpc} waits for the incoming process data of the \glspl{icps}, performs the tasks in the assigned \glspl{vpf} and sends the control data back. The special feature here is that all \glspl{vpc}, both inactive and active, receive and process the data values, but only the active unit transmits output values to the actuators to ensure that the inactive \gls{vpc} can take over this task in case of a failure of the active %\gls{vpc}
one. This requires a precise time and state synchronization of these components.

\subsection{Replacement (Scenario 3)}
This scenario describes how to ensure interruption-free operation of the entire system in case of a failure of one of the \glspl{vpc}%, either the active or the inactive one
. Since the simultaneous failure of the active and all inactive \glspl{vpc} is not considered, two cases are covered. In the first case, the failure of one of the inactive \glspl{vpc} is considered. A malfunction of an inactive \gls{vpc} or an interruption of the communication link due to a failure of the %communication 
infrastructure would be detected by the active %\gls{vpc} 
one by the missing response to the synchronization message. In this situation, the active \gls{vpc} requests the \gls{vpcmo} to create a new backup facility %\gls{vpc} 
to ensure the required reliability. If a suitable \gls{ir} is available, the \gls{vpcmo} will promote it to an inactive \gls{vpc}. If the active \gls{vpc} fails, the inactive \glspl{vpc} recognize this by the absence of the synchronization message. In this case, a predefined inactive \gls{vpc} will be automatically promoted to be the active %\gls{vpc}
one. In the next step, the now active \gls{vpc} instructs the \gls{vpcmo} to replace the inactive %\gls{vpc}
one, similar to the failure of the inactive %\gls{vpc}
unit. If only a failure of the communication link between the two %\glspl{vpc} 
entities is responsible for the missing message, the inactive \gls{vpc} incorrectly assumes the failure of the active \gls{vpc} and incorrectly promotes itself. If the device is completely disconnected from the network due to infrastructure failure, this condition will not affect the system. However, in all other cases, the \gls{vpcmo} would be informed of the problem by a double status message from both active \glspl{vpc} and disable one of them. %the two \gls{vpc} would be disabled.

\subsection{Reconfiguration and Redeployment (Scenario 4)}
If a reconfiguration or redeployment is triggered, the \gls{vpcmo} searches for available \glspl{ir}, selects the two most appropriate ones, and promotes them to $VPC_3$ and $VPC_4$. Once they have completed their startup processes, the \gls{vpcmo} triggers the handover between the previously active $VPC_1$ and the from now on active $VPC_3$ for a given date. Consequently, these two \glspl{vpc} should be highly time synchronized. Eventually, the obsolete \glspl{vpc} are no longer needed and are released.

%#################################################################
%##=========================================================######
%##---------------------------------------------------------######
\section{Virtualization Strategy}
\label{sec: Virtualization Strategy}
%##=========================================================######
%#################################################################
\begin{figure*}[htbp]
\centerline{\includegraphics[width=0.58\textwidth]{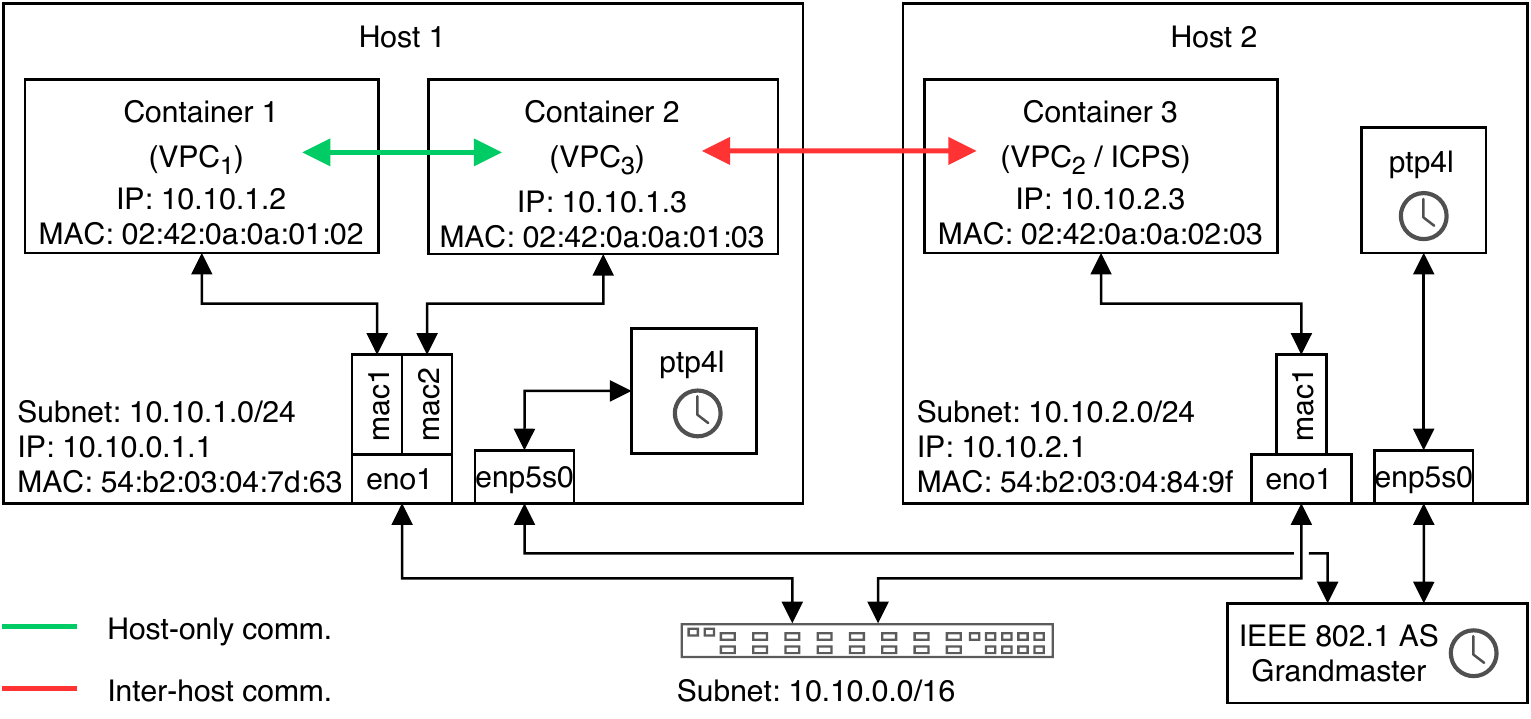}}
	\caption{Testbed configuration for a first evaluation of our virtualization strategy of the proposed architecture}
\label{fig:Testbed configuration}
\end{figure*}

A key task is to determine which virtualization technology can be used to benefit from the proposed architecture. As mentioned above, there are several ongoing activities on this topic, with \cite{7095802} suggesting that virtualization in the \gls{iiot} using containers is more appropriate than \glspl{vm} in terms of memory utilization, latency, and redeployment. However, there are still applications where the use of \glspl{vm} \cite{indin2020} is preferable. Based on the fact that we want to address \gls{rt} applications, including high determinism and low latency, we first map the proposed architecture to container technology. Starting from the results in \cite{indin2020}, where several network configurations were compared, we expect that the macvlan driver is well suited for industrial use. %Major reasons for this assumption are the offered performance that is comparable to bare-metal, an automated deployment using Swarm services, and the fact that each of the containers has its own MAC and IP address. This results in the state, that each container appears as a physical device. 
Major reasons for this assumption are the performance, which is comparable to bare-metal, an automated deployment using Swarm services and the fact that each of the containers has its own MAC and IP address. 
Thus, each container appears as a physical device and has the possibility to use both, \gls{l2} and \gls{l3} communcation. To test the performance of our concept, we use the testbed shown in Fig. \ref{fig:Testbed configuration} as the basis for several tests, where the used hardware is listed in Tab. \ref{tab:hardware}.

\begin{table}[htbp]
\caption{Hardware configurations}
\begin{center}
\begin{tabular*}{\columnwidth}{|c|c|p{0.5\columnwidth}|}
\cline{1-3} %\hline 
\textbf{\textit{Equipment}} & \textbf{\textit{QTY}} & \textbf{\textit{Specification}}\\
\cline{1-3} %\hline 
Mini PC & 2 & Intel i7-8809G, 32 GB DDR4, Intel i210-AT \& i219-LM NICs, Linux 4.19.103-rt42 \\ %Ubuntu 18.04 LTS 64-bit, 
%& & Processor: Intel Core i7-8809G  \\
%& & Kernel: Linux 4.19.103-rt42  \\
\cline{1-3} 
Network Switch & 1 & 8-Port Ethernet Switch\\
\cline{1-3}%\hline 
TSN Eval. Kit & 1 & RAPID-TSNEK-V0001, IEEE~802.1AS-REV \\%, 802.1Qbv, 802.1Qci, 802.1CB, 802.1Qcc, 802.1Qbu / 802.3br \\
\cline{1-3} %\hline 
\end{tabular*}
\label{tab:hardware}
\end{center}
\end{table}

\begin{figure}[htbp]
\resizebox{\columnwidth}{!}{%
\input{figures/e2e2h_final.tikz}
}
	\caption{E2E latency for applications and containers across hosts.}
\label{fig:2host}
\end{figure}
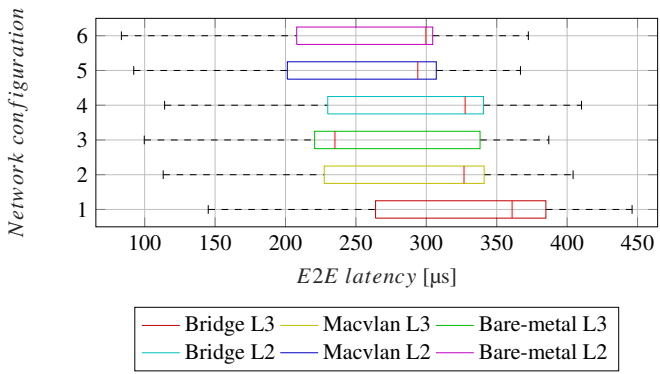

As we investigate first of all if time critical applications can be fulfilled by the applied virtualization, we will concentrate on the communication between \gls{vpc} and \gls{icps} and between \glspl{vpc}. Due to the fact that there are scenarios where it is preferable that both \glspl{vpc} are on the same server to avoid an additional network, and scenarios where \glspl{vpc} must be distributed across multiple servers, the testbed includes two hosts connected by an 8-port network switch. In addition, Host 1 runs two containers ($VPC_1$ and $VPC_3$) to simulate Scenario 4, and Host 2 runs a third container, which can be either $VPC_2$ (Scenario 3) or an \gls{icps} (Scenario 2). To use automatic IP assignment by Docker and to avoid address conflicts, each host has its own /24 subnet. With this configuration, each host is capable of running 253 containers. Since the \gls{e2e} latency of communication between the components is to be determined so that performance and determinism of the framework can be evaluated, both hosts must be time-synchronized. Therefore each host is connected to a shared IEEE 802.1AS grandmaster and runs the ptp4l service, which is part of Linux PTP. Linux PTP\footnote{Further information: http://linuxptp.sourceforge.net/} is a free and open source %\gls{gptp} 
gPTP implementation that complies with the IEEE 802.1AS standard. By using this implementation, a synchronization accuracy of \flq 1µs between both hosts can be guaranteed. To validate the better performance of the macvlan network driver, we compare it to the bridge driver, i.e. the standard configuration of Docker containers, and bare-metal. In addition, the use of messages based directly on the MAC layer (\gls{l2}) has several advantages over the use of the IP layer (\gls{l3}), such as reduced packet size and faster packet processing. For this reason, the \gls{e2e} latency measurements are performed for both \gls{l2} and \gls{l3} communications. The results of the tests are shown in Fig. \ref{fig:1host} and Fig. \ref{fig:2host}.  

 \begin{figure}[htbp]
\resizebox{\columnwidth}{!}{%
\input{figures/e2e1h_final.tikz}
}
	\caption{E2E latency for applications and containers within a single host.}
\label{fig:1host}
\end{figure}
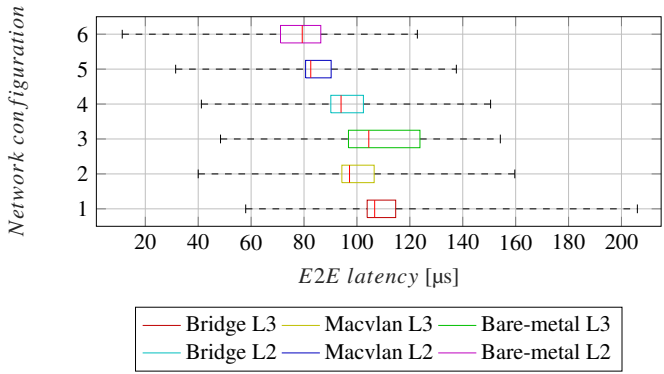

%Especially in Fig. \ref{fig:2host} several conclusions can be taken. First, the the maximum values of the \gls{e2e} latency for bare-metal and macvlan differ only slightly for \gls{l2} and \gls{l3}. Secondly, the transmission of data packets at \gls{l2} saves $\approx$ 40µs. The use of macvlan networks is about 30µs faster on \gls{l2} and 40µs on \gls{l3} compared to bridge network driver. This means, that transmitting messages on \gls{l2} using macvlan network driver can save $\approx$ 60 - 70µs one way, compared with bridge mode that is the default driver for Docker networks. Since most time critical use cases, such as closed loop motion control require two way message exchange in one loop, the time saving is already \frq 120µs. This means, that for time sensitive communication between \glspl{vpc}, such as the state synchronization, the use of \gls{l2} and macvlan network driver offer benefits in puncto performance.
Several conclusions can be derived from both figures. First, the maximum values of the \gls{e2e} latency for bare-metal and macvlan differ only slightly. Furthermore, the \gls{e2e} latency of packet transmission at \gls{l2} is \mbox{$\approx$40µs} lower, compared to \gls{l3}. In addition, a macvlan network is \mbox{$\approx$30µs} faster at \gls{l2} and \mbox{$\approx$40µs} faster at \gls{l3} compared to a bridge network. This means that transferring messages on \gls{l2} using the macvlan network driver in one direction can save \mbox{$\approx$60 - 70µs} compared to the standard Docker network driver. Since most time-critical use cases, such as closed loop motion control, require bilateral message exchange within a single loop, the time saving is already more than 120µs. This leads to the conclusion that for time-critical communication between \glspl{vpc}, such as state synchronization, where no routing is required, the use of \gls{l2} macvlan networks is suitable. %offers advantages compared to other configurations.
%#################################################################
%##=========================================================######
%##---------------------------------------------------------######
\section{Conclusion}%
\label{sec:Conclusion}
%##=========================================================######
%#################################################################
The introduction of virtualization concepts makes it possible to redeploy and reconfigure future industrial automation systems. In this paper we presented a novel architecture that addresses these challenges. Based on the modelling rules of the 4+1 model, we built the architecture from four views and four scenarios. In addition, first insights into the intended virtualization and realization of the architecture are given and initial performance benchmarks for the planned virtualization are presented.  

%\section*{Acknowledgment}%
%##=========================================================######
%#################################################################
%This research was supported by the German Federal Ministry of Education and Research (BMBF) within the project TACNET 4.0 %\gls{tacnet4.0} 
%under grant number 16KIS0712K. The responsibility for this publication lies with the authors.

%\nocite
\nobalance
\printbibliography%
\nobalance
\nl
\nobalance
%#################################################################
%##=========================================================######
%##---------------------------------------------------------######
% - - - Tempory active while working state. Deactivated before \begin{document}
%##=========================================================######
%#################################################################
\TempDisplayPreparation
\end{document}

%% file: organization/preamble.tex
\PassOptionsToPackage{usenames,dvipsnames,svgnames,x11names,table,prologue}{xcolor}%
\PassOptionsToPackage{hyphens}{url}%
\PassOptionsToPackage{nomessages}{fp}%
\ifCLASSINFOpdf
  \usepackage[pdftex]{graphicx}
\else
  \usepackage[dvips]{graphicx}
\fi
\ifCLASSOPTIONcompsoc
   \usepackage[caption=false,font=normalsize,labelfont=sf,textfont=sf]{subfig}
\else
   \usepackage[caption=false,font=footnotesize]{subfig}
\fi

\usepackage{stfloats}
\usepackage[english]{babel}%
\usepackage[utf8]{inputenc}%
\usepackage[babel,style=english]{csquotes}%
\usepackage{hyphsubst}%
\usepackage[%
	activate={true,%
	nocompatibility},%
	final,%
	tracking=true,%
	kerning=true,%
	spacing=true,%
	factor=1100,%
	stretch=10,%
	shrink=10%
]{microtype}%
\usepackage{setspace}%
\usepackage{xcolor}%
\definecolor{todonotecol}{RGB}{250,0,0}%
\usepackage{xparse}
\usepackage[%
	colorlinks=false,%
	urlcolor=black,%
	linkcolor=black,%
	citecolor=black,%
	filecolor=black,%
	breaklinks,%
	]{hyperref}%
\usepackage{url}%
\usepackage[%
	acronym,%
	nopostdot,%
	seeautonumberlist,%
	shortcuts,%
	section=chapter,%
	toc,%
]{glossaries}%
\loadglsentries{./supply/glossaries.tex}\glsdisablehyper% Disables Hyperlinks
\usepackage[%
	backend=biber,%
	style=ieee,%
	isbn=false,%
	hyperref=true,%
	maxbibnames=99,%
	sorting=none,%
	natbib=true,%
	language=english,%
	defernumbers=true,%
	]{biblatex}%
\DeclareFieldFormat{sentencecase}{\csname bbx@colon@search\endcsname#1}% Therewith capital letters stay capitals, when using 'style=ieee'

\usepackage{balance}
\usepackage{nameref}
\usepackage{soul}
\usepackage{flushend}
\usepackage{textcomp}
\usepackage{calc}
\usepackage{xkeyval}
\usepackage{multirow}
\usepackage{tabulary}% Nearly the same as 'tabularx', bit more extensive
\usepackage{makecell}% http://ctan.org/pkg/makecell
\usepackage{pgfplots}
\usepackage{tikz}
\usepackage{textcomp} 
\usepackage{verbatim}

%% file: supply/glossaries.tex
%________________________________________________________________________
%------------------------------------------------------------------------
%					Custom Key Definitions
%/\/\/\/\/\/\/\/\/\/\/\/\/\/\/\/\/\/\/\/\/\/\/\/\/\/\/\/\/\/\/\/\/\/\/\/\%
\glsaddkey*{url}%Key Name
% {\glsentrytext{\glslabel}deutsch}%Default Value
{}%Default Value
{\glsentryurl}%Use analog \glsentrytext
{\Glsentryurl}%Use analog \Glsentrytext
{\glsurl}%Use analog \glstext
{\Glsurl}%Use analog \Glstext
{\GLSurl}%Use analog \GLStext%
\glsaddkey*{longpltwo}%Key Name
% {\glsentrytext{\glslabel}deutsch}%Default Value
{}%Default Value
{\glsentrylongpltwo}%Use analog \glsentrytext
{\Glsentrylongpltwo}%Use analog \Glsentrytext
{\glslongpltwo}%Use analog \glstext
{\Glslongpltwo}%Use analog \Glstext
{\GLSlongpltwo}%Use analog \GLStext%
%/\/\/\/\/\/\/\/\/\/\/\/\/\/\/\/\/\/\/\/\/\/\/\/\/\/\/\/\/\/\/\/\/\/\/\/\
%					Custom Key Definitions fertig
%------------------------------------------------------------------------
%________________________________________________________________________
%
%
%
%
%
%________________________________________________________________________
%------------------------------------------------------------------------
%						Glossar
%/\/\/\/\/\/\/\/\/\/\/\/\/\/\/\/\/\/\/\/\/\/\/\/\/\/\/\/\/\/\/\/\/\/\/\/\
\newglossaryentry{}{%
	name={},%
	plural={},%
	description={},%
}%
%
%/\/\/\/\/\/\/\/\/\/\/\/\/\/\/\/\/\/\/\/\/\/\/\/\/\/\/\/\/\/\/\/\/\/\/\/\
%						Glossar fertig
%------------------------------------------------------------------------
%________________________________________________________________________
%
%
%
%
%
%________________________________________________________________________
%------------------------------------------------------------------------
%						Akronyme (Abkürzungen)
%/\/\/\/\/\/\/\/\/\/\/\/\/\/\/\/\/\/\/\/\/\/\/\/\/\/\/\/\/\/\/\/\/\/\/\/\
\newacronym{it}{IT}{information technology}
\newacronym{ot}{OT}{operational technology}
\newacronym{opcua}{OPC UA}{Open Platform Communications Unified Architecture}
\newacronym{tsn}{TSN}{time-sensitive networking}
\newacronym{scada}{SCADA}{supervisory control and data acquisition}
\newacronym{vpc}{VPC}{Virtual Process Controller}
\newacronym{plc}{PLC}{programmable logic controller}
\newacronym{udp}{UDP}{User Datagram Protocol}
\newacronym{pubsub}{Pub/Sub}{Publish/Subscribe}
\newacronym{vm}{VM}{virtual machine}
%\newacronym{IEEE}[IEEE]{Institute of Electrical and Electronics Engineers}
\newacronym{pc}{PC}{personal computer}
\newacronym{ptp}{PTP}{Precision Time Protocol}
\newacronym{gptp}{gPTP}{generalized Precision Time Protocol}
\newacronym{ntp}{NTP}{Network Time Protocol}
\newacronym{splc}{Soft-PLC}{Software-PLC}
\newacronym{rte}{RTE}{Real-Time Ethernet}
\newacronym{osi}{OSI}{Open Systems Interconnection}
\newacronym{mac}{MAC}{Media Access Control}
\newacronym{e2e}{E2E}{end-to-end}
\newacronym{tcp}{TCP}{Transmission Control Protocol}
\newacronym{hmi}{HMI}{Human-Machine Interface}
\newacronym{kpi}{KPI}{Key Performance Indicator}
\newacronym{mes}{MES}{Manufacturing Execution System}
\newacronym{rami4.0}{RAMI 4.0}{Reference Architectural Model Industrie 4.0}
\newacronym{tacnet4.0}{TACNET 4.0}{Tactile Internet 4.0}
\newacronym{3gpp}{3GPP}{3rd Generation Partnership Project}
\newacronym{5gppp}{5GPPP}{5G Infrastructure Public Private Partnership}
\newacronym{itu}{ITU}{International Telecommunication Union}
\newacronym{ngnm}{NGNM}{Next Generation Mobile Networks}
\newacronym{agv}{AGV}{automated guided vehicle}
\newacronym{v2v}{V2V}{vehicle-to-vehicle}
\newacronym{v2x}{V2X}{vehicle-to-infrastructure}
\newacronym{d2x}{D2X}{device-to-x}
\newacronym{qos}{QoS}{quality of service}
\newacronym{noa}{NOA}{NAMUR Open Architecture}
\newacronym{dcs}{DCS}{distributed control system}
\newacronym{m2m}{M2M}{machine-to-machine}
\newacronym{sdn}{SDN}{software-defined networking}
\newacronym{erp}{ERP}{enterprise resource planning}
\newacronym{ip}{IP}{Internet Protocol}
\newacronym{lte}{LTE}{Long Term Evolution}
\newacronym{5g}{5G}{5th generation wireless communication systems}
\newacronym{harq}{HARQ}{Hybrid Automatic Repeat Request}
\newacronym{rat}{RAT}{radio access technology}
\newacronym{mc}{MC}{Multi-Connectivity}
\newacronym{fbmc}{FBMC}{Filter Bank Multicarrier}
\newacronym{gfdm}{GFDM}{Generalized Frequency Division Multiplexing}
\newacronym{sla}{SLA}{service-level agreement}
\newacronym{5qi}{5QI}{5G QoS Indicator}
\newacronym{bmbf}{BMBF}{German Federal Ministry of Education and Research}
\newacronym{nrt}{NRT}{non real-time}
\newacronym{irt}{IRT}{isochronous real-time}
\newacronym{rt}{RT}{real-time}
\newacronym{ar}{AR}{augmented reality}
\newacronym{wlan}{WLAN}{wireless local area network}
\newacronym{cpu}{CPU}{central processing unit}
\newacronym{ram}{RAM}{random-access memory}
\newacronym{ue}{UE}{user equipment}
\newacronym{ran}{RAN}{radio access network}
\newacronym{bs}{BS}{base station}
\newacronym{cn}{CN}{core network}
\newacronym{epc}{EPC}{evolved packet core}
\newacronym{mec}{MEC}{mobile edge cloud}
\newacronym{rtt}{RTT}{round-trip time}
\newacronym{vpn}{VPN}{virtual private network}
\newacronym{vpf}{VPF}{Virtual Process Control Function}
\newacronym{vdi}{VDI}{Association of German Engineers}
\newacronym{ipc}{IPC}{industrial pc}
\newacronym{rtos}{RTOS}{real-time operating system}
\newacronym{vpcmo}{M\&O}{VPC Management \& Orchestration}
\newacronym{i/o}{I/O}{input /output}
\newacronym{fbd}{FBD}{Function Block Diagram}
\newacronym{ldd}{LDD}{Ladder Logic Diagram}
\newacronym{sfc}{SFC}{Sequential Function Chart}
\newacronym{il}{IL}{Instruction List}
\newacronym{st}{ST}{Structured Text}
\newacronym{os}{OS}{operating system}
\newacronym{ir}{IR}{Inactive Resource}
\newacronym{msc}{MSC}{message sequence chart}
\newacronym{icps}{ICPS}{industrial cyber-physical system}
\newacronym{cps}{CPS}{cyber-physical systems}
\newacronym{mtd}{MTD}{Moving Target Defense}
\newacronym{iot}{IoT}{Internet of Things}
\newacronym{iiot}{IIoT}{industrial Internet of Things}
\newacronym{iiot2}{IIoT}{Industrial IoT}
\newacronym{k8s}{K8s}{Kubernetes}
\newacronym{rkt}{rkt}{CoreOS Rocket}
\newacronym{dcos}{DC/OS}{Distributed Cloud Operating System}
\newacronym{ml}{ML}{machine learning}
\newacronym{ai}{AI}{artificial intelligence}
\newacronym{ict}{ICT}{industrial communication technology}
\newacronym{kvm}{KVM}{kernel-based virtual machine}
\newacronym{l2}{L2}{layer 2}
\newacronym{l3}{L3}{layer 3}

%% file: organization/makros.tex
\newcommand{\nl}{\par\noindent} % Abk. für NewLine
\newcommand{\mytilde}{{\raise.17ex\hbox{$\scriptstyle\mathtt{\sim}$}}}

%% file: organization/settings.tex
\newlength\textheighttemp%
\newlength\textwidthtemp%
\newlength\textheightstd%
\newlength\textwidthstd%
\newlength\textheightold%
\newlength\textwidthold%
\newlength\tempheight%
\newlength\tempwidth%
\SetProtrusion{encoding={*},family={bch},series={*},size={6,7}}
              {1={ ,750},2={ ,500},3={ ,500},4={ ,500},5={ ,500},
               6={ ,500},7={ ,600},8={ ,500},9={ ,500},0={ ,500}}
\SetExtraKerning[unit=space]
    {encoding={*}, family={bch}, series={*}, size={footnotesize,small,normalsize}}
    {\textendash={400,400}, % en-dash, add more space around it
     "28={ ,150}, % left bracket, add space from right
     "29={150, }, % right bracket, add space from left
     \textquotedblleft={ ,150}, % left quotation mark, space from right
     \textquotedblright={150, }} % right quotation mark, space from left
\SetTracking{encoding={*}, shape=sc}{40}

%% file: organization/IEEE_authors-long.tex
\author{%
\IEEEauthorblockN{%
    Michael Gundall\IEEEauthorrefmark{1}, %
    Calvin Glas\IEEEauthorrefmark{1}, %
    and Hans D. Schotten\IEEEauthorrefmark{1}\IEEEauthorrefmark{2} %
    \\%
  %  FirstName1 Lastname1\IEEEauthorrefmark{3} and %
   % FirstName2 Lastname2\IEEEauthorrefmark{4}%
}%
\IEEEauthorblockA{%
    \IEEEauthorrefmark{1}German Research Center for Artificial Intelligence GmbH (DFKI), \\ Kaiserslautern, Germany \\% 
  %  \IEEEauthorrefmark{2} Department of Computer Science, Technische Universität Kaiserslautern, \\ Kaiserslautern, Germany \\%
    \IEEEauthorrefmark{2}Department of Electrical and Computer Engineering, Technische Universität Kaiserslautern, \\ Kaiserslautern, Germany %
%     Trippstadter Str. 122\\%
%     67663 Kaiserslautern\\%
	\\%
  %  \IEEEauthorrefmark{3}Institute1, %
  %  Some Subtitle 1 %
%     Optional Address, Germany%
 %   \\%
	%\IEEEauthorrefmark{4}Corporation2, %
  %  Some Subtitle2, %
  %  Some more Subt2 %
%     Optional Address
  %  \\%
    Email: %
        \{michael.gundall, calvin.glas, hans\_dieter.schotten%
        \}@dfki.de%, %
        \\%
    %    c\_glas15@informatik.uni-kl.de
     %   \\
      %  \IEEEauthorrefmark{3}Mail1@domain1.de, %
      %  \IEEEauthorrefmark{4}Mail2@domain2.com
}%
}%


%Telephone: (800) 555--1212, Fax: (888) 555--1212}

%% file: figures/e2e1h_final.tikz
\definecolor{mycolor1}{rgb}{0.75000,0.75000,0.00000}%
\definecolor{mycolor2}{rgb}{0.00000,0.75000,0.75000}%
\definecolor{mycolor3}{rgb}{0.75000,0.00000,0.75000}%
\begin{tikzpicture}

\begin{axis}[%
width=3.5in,
height=1.3in,
scale only axis,
unbounded coords=jump,
xmin=1.5,
xmax=215,
xlabel style={font=\color{white!15!black}},
xlabel={$E2E~latency$ [µs]},
ymin=0.5,
ymax=6.5,
ytick={1,2,3,4,5,6},
ylabel style={font=\color{white!15!black}},
ylabel={$Network~configuration$},
axis background/.style={fill=white},
xmajorgrids,
ymajorgrids,
legend style={at={(1.75 in, -0.7 in)}, anchor=center, legend columns=3 draw=black}
]
\addplot [color=black, dashed, forget plot]
  table[row sep=crcr]{%
114.65	1\\
206.012	1\\
};
\addplot [color=black, dashed, forget plot]
  table[row sep=crcr]{%
106.51	2\\
159.677	2\\
};
\addplot [color=black, dashed, forget plot]
  table[row sep=crcr]{%
123.822	3\\
154.205	3\\
};
\addplot [color=black, dashed, forget plot]
  table[row sep=crcr]{%
102.446	4\\
150.525	4\\
};
\addplot [color=black, dashed, forget plot]
  table[row sep=crcr]{%
90.254	5\\
137.581	5\\
};
\addplot [color=black, dashed, forget plot]
  table[row sep=crcr]{%
86.335	6\\
122.845	6\\
};
\addplot [color=black, dashed, forget plot]
  table[row sep=crcr]{%
57.951	1\\
103.87025	1\\
};
\addplot [color=black, dashed, forget plot]
  table[row sep=crcr]{%
39.999	2\\
94.303	2\\
};
\addplot [color=black, dashed, forget plot]
  table[row sep=crcr]{%
48.431	3\\
96.798	3\\
};
\addplot [color=black, dashed, forget plot]
  table[row sep=crcr]{%
41.199	4\\
90.126	4\\
};
\addplot [color=black, dashed, forget plot]
  table[row sep=crcr]{%
31.471	5\\
80.59475	5\\
};
\addplot [color=black, dashed, forget plot]
  table[row sep=crcr]{%
11.28	6\\
71.10625	6\\
};
\addplot [color=black, forget plot]
  table[row sep=crcr]{%
206.012	0.875\\
206.012	1.125\\
};
\addplot [color=black, forget plot]
  table[row sep=crcr]{%
159.677	1.875\\
159.677	2.125\\
};
\addplot [color=black, forget plot]
  table[row sep=crcr]{%
154.205	2.875\\
154.205	3.125\\
};
\addplot [color=black, forget plot]
  table[row sep=crcr]{%
150.525	3.875\\
150.525	4.125\\
};
\addplot [color=black, forget plot]
  table[row sep=crcr]{%
137.581	4.875\\
137.581	5.125\\
};
\addplot [color=black, forget plot]
  table[row sep=crcr]{%
122.845	5.875\\
122.845	6.125\\
};
\addplot [color=black, forget plot]
  table[row sep=crcr]{%
57.951	0.875\\
57.951	1.125\\
};
\addplot [color=black, forget plot]
  table[row sep=crcr]{%
39.999	1.875\\
39.999	2.125\\
};
\addplot [color=black, forget plot]
  table[row sep=crcr]{%
48.431	2.875\\
48.431	3.125\\
};
\addplot [color=black, forget plot]
  table[row sep=crcr]{%
41.199	3.875\\
41.199	4.125\\
};
\addplot [color=black, forget plot]
  table[row sep=crcr]{%
31.471	4.875\\
31.471	5.125\\
};
\addplot [color=black, forget plot]
  table[row sep=crcr]{%
11.28	5.875\\
11.28	6.125\\
};
\addplot [color=black!25!red]
  table[row sep=crcr]{%
103.87025	0.75\\
114.65	0.75\\
114.65	1.25\\
103.87025	1.25\\
103.87025	0.75\\
};
\addlegendentry{Bridge L3}

\addplot [color=mycolor1]
  table[row sep=crcr]{%
94.303	1.75\\
106.51	1.75\\
106.51	2.25\\
94.303	2.25\\
94.303	1.75\\
};
\addlegendentry{Macvlan L3}

\addplot [color=black!25!green]
  table[row sep=crcr]{%
96.798	2.75\\
123.822	2.75\\
123.822	3.25\\
96.798	3.25\\
96.798	2.75\\
};
\addlegendentry{Bare-metal L3}

\addplot [color=mycolor2]
  table[row sep=crcr]{%
90.126	3.75\\
102.446	3.75\\
102.446	4.25\\
90.126	4.25\\
90.126	3.75\\
};
\addlegendentry{Bridge L2}

\addplot [color=black!25!blue]
  table[row sep=crcr]{%
80.59475	4.75\\
90.254	4.75\\
90.254	5.25\\
80.59475	5.25\\
80.59475	4.75\\
};
\addlegendentry{Macvlan L2}

\addplot [color=mycolor3]
  table[row sep=crcr]{%
71.10625	5.75\\
86.335	5.75\\
86.335	6.25\\
71.10625	6.25\\
71.10625	5.75\\
};
\addlegendentry{Bare-metal L2}

\addplot [color=red, forget plot]
  table[row sep=crcr]{%
106.702	0.75\\
106.702	1.25\\
};
\addplot [color=red, forget plot]
  table[row sep=crcr]{%
97.199	1.75\\
97.199	2.25\\
};
\addplot [color=red, forget plot]
  table[row sep=crcr]{%
104.478	2.75\\
104.478	3.25\\
};
\addplot [color=red, forget plot]
  table[row sep=crcr]{%
93.998	3.75\\
93.998	4.25\\
};
\addplot [color=red, forget plot]
  table[row sep=crcr]{%
82.542	4.75\\
82.542	5.25\\
};
\addplot [color=red, forget plot]
  table[row sep=crcr]{%
79.326	5.75\\
79.326	6.25\\
};
\addplot [color=black, draw=none, mark size=1.5pt, mark=x, mark options={solid, black!25!red}, forget plot]
  table[row sep=crcr]{%
nan	nan\\
};
\addplot [color=black, draw=none, mark size=1.5pt, mark=x, mark options={solid, mycolor1}, forget plot]
  table[row sep=crcr]{%
nan	nan\\
};
\addplot [color=black, draw=none, mark size=1.5pt, mark=x, mark options={solid, black!25!green}, forget plot]
  table[row sep=crcr]{%
nan	nan\\
};
\addplot [color=black, draw=none, mark size=1.5pt, mark=x, mark options={solid, mycolor2}, forget plot]
  table[row sep=crcr]{%
nan	nan\\
};
\addplot [color=black, draw=none, mark size=1.5pt, mark=x, mark options={solid, black!25!blue}, forget plot]
  table[row sep=crcr]{%
nan	nan\\
};
\addplot [color=black, draw=none, mark size=1.5pt, mark=x, mark options={solid, mycolor3}, forget plot]
  table[row sep=crcr]{%
nan	nan\\
};
\end{axis}
\end{tikzpicture}%

%% file: figures/e2e2h_final.tikz
\definecolor{mycolor1}{rgb}{0.75000,0.75000,0.00000}%
\definecolor{mycolor2}{rgb}{0.00000,0.75000,0.75000}%
\definecolor{mycolor3}{rgb}{0.75000,0.00000,0.75000}%
\begin{tikzpicture}

\begin{axis}[%
width=3.5in,
height=1.3in,
scale only axis,
unbounded coords=jump,
xmin=65.34205,
xmax=464.26695,
xlabel style={font=\color{white!15!black}},
xlabel={$E2E~latency$ [µs]},
ymin=0.5,
ymax=6.5,
ytick={1,2,3,4,5,6},
ylabel style={font=\color{white!15!black}},
ylabel={$Network~configuration$},
axis background/.style={fill=white},
xmajorgrids,
ymajorgrids,
legend style={at={(1.75 in, -0.7 in)}, anchor=center, legend columns=3 draw=black}
]
\addplot [color=black, dashed, forget plot]
  table[row sep=crcr]{%
384.9125	1\\
446.134	1\\
};
\addplot [color=black, dashed, forget plot]
  table[row sep=crcr]{%
341.0745	2\\
404.237	2\\
};
\addplot [color=black, dashed, forget plot]
  table[row sep=crcr]{%
338.1675	3\\
386.996	3\\
};
\addplot [color=black, dashed, forget plot]
  table[row sep=crcr]{%
340.60925	4\\
410.202	4\\
};
\addplot [color=black, dashed, forget plot]
  table[row sep=crcr]{%
307.078	5\\
366.859	5\\
};
\addplot [color=black, dashed, forget plot]
  table[row sep=crcr]{%
304.5485	6\\
372.454	6\\
};
\addplot [color=black, dashed, forget plot]
  table[row sep=crcr]{%
145.147	1\\
263.97225	1\\
};
\addplot [color=black, dashed, forget plot]
  table[row sep=crcr]{%
113.161	2\\
227.51225	2\\
};
\addplot [color=black, dashed, forget plot]
  table[row sep=crcr]{%
99.719	3\\
220.692	3\\
};
\addplot [color=black, dashed, forget plot]
  table[row sep=crcr]{%
114.173	4\\
230.0075	4\\
};
\addplot [color=black, dashed, forget plot]
  table[row sep=crcr]{%
92.238	5\\
201.30675	5\\
};
\addplot [color=black, dashed, forget plot]
  table[row sep=crcr]{%
83.475	6\\
207.9695	6\\
};
\addplot [color=black, forget plot]
  table[row sep=crcr]{%
446.134	0.875\\
446.134	1.125\\
};
\addplot [color=black, forget plot]
  table[row sep=crcr]{%
404.237	1.875\\
404.237	2.125\\
};
\addplot [color=black, forget plot]
  table[row sep=crcr]{%
386.996	2.875\\
386.996	3.125\\
};
\addplot [color=black, forget plot]
  table[row sep=crcr]{%
410.202	3.875\\
410.202	4.125\\
};
\addplot [color=black, forget plot]
  table[row sep=crcr]{%
366.859	4.875\\
366.859	5.125\\
};
\addplot [color=black, forget plot]
  table[row sep=crcr]{%
372.454	5.875\\
372.454	6.125\\
};
\addplot [color=black, forget plot]
  table[row sep=crcr]{%
145.147	0.875\\
145.147	1.125\\
};
\addplot [color=black, forget plot]
  table[row sep=crcr]{%
113.161	1.875\\
113.161	2.125\\
};
\addplot [color=black, forget plot]
  table[row sep=crcr]{%
99.719	2.875\\
99.719	3.125\\
};
\addplot [color=black, forget plot]
  table[row sep=crcr]{%
114.173	3.875\\
114.173	4.125\\
};
\addplot [color=black, forget plot]
  table[row sep=crcr]{%
92.238	4.875\\
92.238	5.125\\
};
\addplot [color=black, forget plot]
  table[row sep=crcr]{%
83.475	5.875\\
83.475	6.125\\
};
\addplot [color=black!25!red]
  table[row sep=crcr]{%
263.97225	0.75\\
384.9125	0.75\\
384.9125	1.25\\
263.97225	1.25\\
263.97225	0.75\\
};
\addlegendentry{Bridge L3}

\addplot [color=mycolor1]
  table[row sep=crcr]{%
227.51225	1.75\\
341.0745	1.75\\
341.0745	2.25\\
227.51225	2.25\\
227.51225	1.75\\
};
\addlegendentry{Macvlan L3}

\addplot [color=black!25!green]
  table[row sep=crcr]{%
220.692	2.75\\
338.1675	2.75\\
338.1675	3.25\\
220.692	3.25\\
220.692	2.75\\
};
\addlegendentry{Bare-metal L3}

\addplot [color=mycolor2]
  table[row sep=crcr]{%
230.0075	3.75\\
340.60925	3.75\\
340.60925	4.25\\
230.0075	4.25\\
230.0075	3.75\\
};
\addlegendentry{Bridge L2}

\addplot [color=black!25!blue]
  table[row sep=crcr]{%
201.30675	4.75\\
307.078	4.75\\
307.078	5.25\\
201.30675	5.25\\
201.30675	4.75\\
};
\addlegendentry{Macvlan L2}

\addplot [color=mycolor3]
  table[row sep=crcr]{%
207.9695	5.75\\
304.5485	5.75\\
304.5485	6.25\\
207.9695	6.25\\
207.9695	5.75\\
};
\addlegendentry{Bare-metal L2}

\addplot [color=red, forget plot]
  table[row sep=crcr]{%
360.929	0.75\\
360.929	1.25\\
};
\addplot [color=red, forget plot]
  table[row sep=crcr]{%
326.772	1.75\\
326.772	2.25\\
};
\addplot [color=red, forget plot]
  table[row sep=crcr]{%
235.121	2.75\\
235.121	3.25\\
};
\addplot [color=red, forget plot]
  table[row sep=crcr]{%
327.565	3.75\\
327.565	4.25\\
};
\addplot [color=red, forget plot]
  table[row sep=crcr]{%
293.964	4.75\\
293.964	5.25\\
};
\addplot [color=red, forget plot]
  table[row sep=crcr]{%
299.704	5.75\\
299.704	6.25\\
};
\addplot [color=black, draw=none, mark size=1.5pt, mark=x, mark options={solid, black!25!red}, forget plot]
  table[row sep=crcr]{%
nan	nan\\
};
\addplot [color=black, draw=none, mark size=1.5pt, mark=x, mark options={solid, mycolor1}, forget plot]
  table[row sep=crcr]{%
nan	nan\\
};
\addplot [color=black, draw=none, mark size=1.5pt, mark=x, mark options={solid, black!25!green}, forget plot]
  table[row sep=crcr]{%
nan	nan\\
};
\addplot [color=black, draw=none, mark size=1.5pt, mark=x, mark options={solid, mycolor2}, forget plot]
  table[row sep=crcr]{%
nan	nan\\
};
\addplot [color=black, draw=none, mark size=1.5pt, mark=x, mark options={solid, black!25!blue}, forget plot]
  table[row sep=crcr]{%
nan	nan\\
};
\addplot [color=black, draw=none, mark size=1.5pt, mark=x, mark options={solid, mycolor3}, forget plot]
  table[row sep=crcr]{%
nan	nan\\
};
\end{axis}
\end{tikzpicture}%